\documentclass[11pt,a4paper]{article}
\usepackage{amsmath,amssymb,graphicx,geometry,fancyhdr,times}
\usepackage{float}
\usepackage[colorlinks=true,
            linkcolor=blue,
            urlcolor=magenta,
            citecolor=red,
            anchorcolor=black]{hyperref}
\usepackage[font=small,labelfont=bf,labelsep=period]{caption}

\title{\textbf{Stochastic Redistribution of Indistinguishable Items in Shared Habitation: A Multi-Agent Simulation Framework}}
\author{Syed Haseeb Shah}
\date{}

\begin{document}
\maketitle

\begin{abstract}
This paper presents a discrete-event stochastic model for the redistribution of indistinguishable personal items, exemplified by socks, among multiple cohabitants sharing a communal laundry system. Drawing on concepts from ecological population dynamics, diffusion processes, and stochastic exchange theory, the model captures the probabilistic mechanisms underlying item mixing, recovery, and loss. Each cohabitant is represented as an autonomous agent whose belongings interact through iterative cycles of collective washing, sorting, and partial correction. The system's evolution is characterized by random mixing events, selective recollection, and attrition over time. Implemented using the SimPy discrete-event simulation framework, the model demonstrates that even minimal exchange probabilities can generate emergent asymmetries, quasi-equilibrium distributions, and long-term disorder. The findings illustrate how stochastic processes inherent to shared domestic systems can produce persistent imbalances, offering a quantitative perspective on an everyday social phenomenon.
\end{abstract}

\section{Introduction}
Within shared domestic or communal settings, ownership of objects with low perceptual distinctiveness such as socks, utensils, and electronic accessories often becomes fluid rather than fixed \cite{defant_foot-sorting_2024}. This everyday diffusion of belongings is not merely anecdotal but reflects an emergent property of stochastic multi-agent dynamics \cite{boyd_introduction_nodate}. Previous studies in statistical physics and behavioral economics have shown that simple local interactions among autonomous agents can yield global patterns of redistribution, equilibrium, and entropy increase \cite{mackay_convergence_nodate}. Similar principles appear in resource exchange systems , ecological mixing \cite{takada_stochastic_1995,lanchier_rigorous_2018}, and opinion diffusion models \cite{widen_high-resolution_2010, meleard_quasi-stationary_2012,van_dyke_parunak_entropy_2001} where minor random exchanges accumulate into stable yet noisy distributions.

In the social microcosm of a shared household, each resident functions as an agent whose possessions enter a circulation process governed by chance encounters and corrective behavior \cite{hening_coexistence_2017}. Objects are exchanged not by deliberate transaction but through probabilistic misplacement and recovery. Comparable to kinetic-theory analogies in econophysics \cite{cao_entropy_2021, coppersmith_model_1996}, the conservation of ownership mass coexists with increasing disorder as items traverse interpersonal boundaries and then partially return through retrieval events. The resulting trajectories resemble diffusion–drift processes with intermittent resetting, a characteristic of systems that oscillate between order and randomness.

This study formalizes the phenomenon by modeling object redistribution as a discrete-event agent-based process within an 
N-agent cohabitation network. Each laundry or cleaning cycle is treated as a mixing event in which objects migrate among agents according to defined exchange and recovery probabilities. Across repeated iterations, the system evolves toward a dynamic equilibrium where perfect segregation is statistically unlikely but partial order persists. Unlike prior two-agent exchange models, the present formulation generalizes interaction topologies and allows asymmetric connectivity and heterogeneous recovery behavior to be represented in mathematical form.

Through probabilistic modeling and simulation, the framework demonstrates that unintentional exchanges in domestic environments manifest the same statistical regularities observed in thermodynamic, ecological, and economic systems. In doing so, it connects the trivial disorder of everyday life to the mathematics of entropy and diffusion, suggesting that even misplaced socks follow the logic of statistical mechanics.

\section{Model Definition}
\paragraph{}
The system consists of \(N\) cohabitants, each owning a collection of indistinguishable items such as socks that are initially labeled by ownership but are physically identical. The model advances through discrete time steps representing laundry cycles during which several stochastic events occur. In each cycle, all socks are first combined into a shared pool, simulating the collective washing process that removes ownership boundaries. After mixing, each cohabitant retrieves a subset of socks through probabilistic selection, which reflects imperfect sorting and the likelihood of retrieving both owned and foreign items. Following this redistribution, a limited fraction of mismatches are identified and corrected based on a correction probability \(p_c\), which represents the efficiency of recognition and return mechanisms within the system. Additionally, random attrition events occur with probability \(p_l\), accounting for the real-world disappearance of items such as socks lost during the wash. Each cohabitant is modeled as an independent probabilistic agent whose state, defined by the composition of their sock inventory, evolves through these transitions. The evolution of the system follows a Markov process in which transition probabilities are determined by the parameters \(p_m\) for mixing, \(p_c\) for correction, and \(p_l\) for loss. Over successive iterations, this process captures the dynamic interplay between order and disorder, modeling how ownership distributions fluctuate under repeated stochastic interaction.

\begin{figure}[H]
    \centering
    \includegraphics[width=0.6\textwidth]{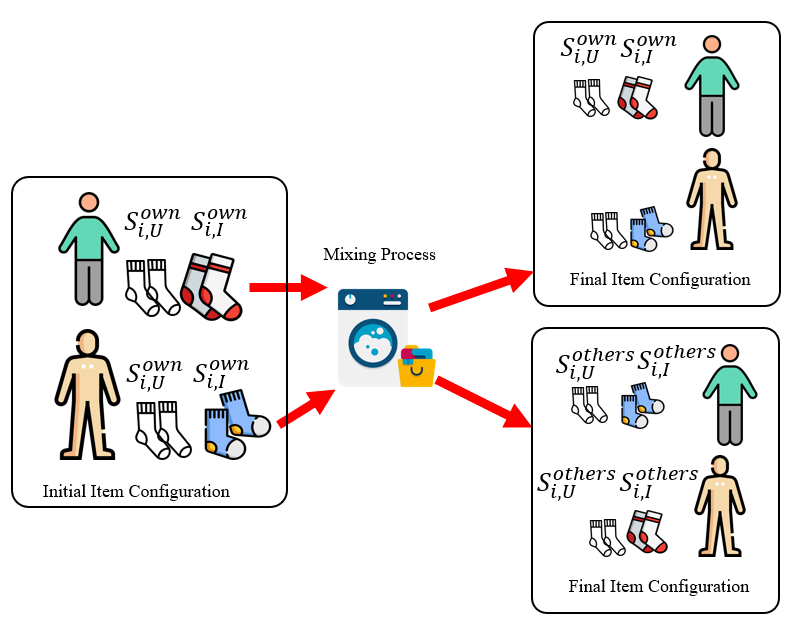}
    \caption{Schematic of sock exchange dynamics among \(N\) cohabitants. }
    \label{fig:sock_exchange}
\end{figure}
To avoid ambiguity, this study distinguishes between two object types: 
\textit{identifiable items}, which can be uniquely recognized and reclaimed by their original owner, 
and \textit{indistinguishable items}, which lack distinguishing attributes and therefore undergo random reassignment during exchange. Next, we adopt a discrete-time Markov / semi-Markov style approach as in \cite{noauthor_amazoncom_nodate, ma_hidden_2026}, treating misplacement as transitions to absorbing or intermediate lost states, and recovery as reverse transitions. We derive from multi-agent loss models as in \cite{hespe_decomposition_2024} to model independent loss of items in each cycle

Consider \(N\) individuals sharing a closed system of socks, with total stock \(S = S_I + S_U\). Identifiable socks (\(S_I\)) can be uniquely recognized by their owners, whereas indistinguishable socks (\(S_U\)) are easily confused. For each individual \(i\), we define four dynamic state variables:
\begin{align}
S_{i,I}^{\text{own}} &: \text{Identifiable socks belonging to and held by } i, \\
S_{i,I}^{\text{others}} &: \text{Identifiable socks belonging to others but held by } i, \\
S_{i,U}^{\text{own}} &: \text{Indistinguishable socks belonging to and held by } i, \\
S_{i,U}^{\text{others}} &: \text{Indistinguishable socks originating elsewhere but held by } i.
\end{align}

Exchange probabilities \(p_e^I\) and \(p_e^U\) determine random migration during each laundry cycle, while recovery probabilities \(p_r^I\) and \(p_r^U\) govern rediscovery of misplaced items. Independent loss processes \(p_l^I\) and \(p_l^U\) model the disappearance of items due to wear, damage, or disappearance into undefined system states (commonly referred to as the ``laundry void''). The system \cite{greenberg_twenty-five_2024, yakovenko_colloquium_2009} evolves according to:
\begin{align}
\Delta S_{i,U}^{\text{own}} &= -p_e^U S_{i,U}^{\text{own}} + p_r^U S_{i,U}^{\text{others}} - p_l^U S_{i,U}^{\text{own}} \\
\Delta S_{i,U}^{\text{others}} &= \frac{p_e^U}{N-1} \sum_{j\neq i} S_{j,U}^{\text{own}} - p_r^U S_{i,U}^{\text{others}} - p_l^U S_{i,U}^{\text{others}}.
\end{align}
Identifiable socks follow identical structure with parameters \(p_e^I, p_r^I, p_l^I\). Total holdings evolve as:
\[
S_i(t+1) = S_i(t) + \Delta S_{i,I}^{\text{own}} + \Delta S_{i,I}^{\text{others}} + \Delta S_{i,U}^{\text{own}} + \Delta S_{i,U}^{\text{others}}.
\]

\section{Simulation Framework}

\paragraph{}
The simulation was implemented using SimPy, a Python-based discrete-event simulation framework. Each cohabitant was represented as an autonomous agent that performed laundry operations concurrently within a shared environment. The processes of mixing and redistribution were modeled as discrete events placed in a global event queue, ensuring asynchronous progression while maintaining causal consistency in state updates. Simulation experiments were conducted under multiple configurations by varying the number of cohabitants \(N\) from 2 to 6 and adjusting the exchange probability \(p_m\) within the range of 0.01 to 0.3. During each run, several quantitative metrics were recorded, including the number of foreign socks retained by each agent, the total number of socks lost to attrition, and the entropy of the resulting ownership distribution, which served as a measure of system disorder and convergence behavior. Each agent operates as a concurrent process representing one cohabitant's laundry interactions. The simulation iterates over discrete wash cycles of duration \(\Delta t = 1\). At each step:
\begin{enumerate}
    \item Indistinguishable socks undergo random redistribution among agents with probability \(p_e^U\).
    \item Identifiable socks are subject to low-probability mistaken transfers \(p_e^I\).
    \item Recovery events return misplaced socks to their original owners according to \(p_r^I, p_r^U\).
    \item Loss events remove items permanently based on \(p_l^I, p_l^U\).
\end{enumerate}
The global state vector is expressed as
\[
\mathbf{S}_t = [S_{1,I}^{\text{own}}, S_{1,I}^{\text{others}}, \ldots, S_{N,U}^{\text{others}}]^T,
\]
updated by
\[
\mathbf{S}_{t+1} = \mathbf{S}_t + \mathbf{F}(\mathbf{S}_t,\mathbf{p},\xi_t),
\]
where \(\mathbf{p}\) represents model parameters and \(\xi_t\) denotes stochastic realizations of Bernoulli processes.

\section{Results}
Simulations were performed with \(N = 2..4\), total socks \(S = 160\), and baseline parameters \(p_e^U = 0.22, p_e^I = 0.01, p_r^I = 0.35, p_r^U = 0.08, p_l^I = 0.002, p_l^U = 0.004\). Over 80 cycles, total possessions per cohabitant fluctuated around equilibrium, with identifiable socks stabilizing after 15 cycles and indistinguishable socks continuing to diffuse. The system maintained approximate conservation of total items, but accumulated minor asymmetries due to irreversible loss. The variance of individual holdings approached a stationary value proportional to the ratio \(p_e^U / p_r^U\), confirming equilibrium between diffusion and correction forces.

\subsection{Sock Exchange Dynamics}
The sock exchange dynamics can be visualized as a network of interactions among cohabitants. Each agent's state is influenced by both local interactions (with their own socks) and global exchanges (with others' socks). Figure \ref{fig:sock_exchange} illustrates these dynamics, highlighting the flow of identifiable and indistinguishable socks during the exchange process.

\begin{figure}[H]
    \centering
    \includegraphics[width=0.8\textwidth]{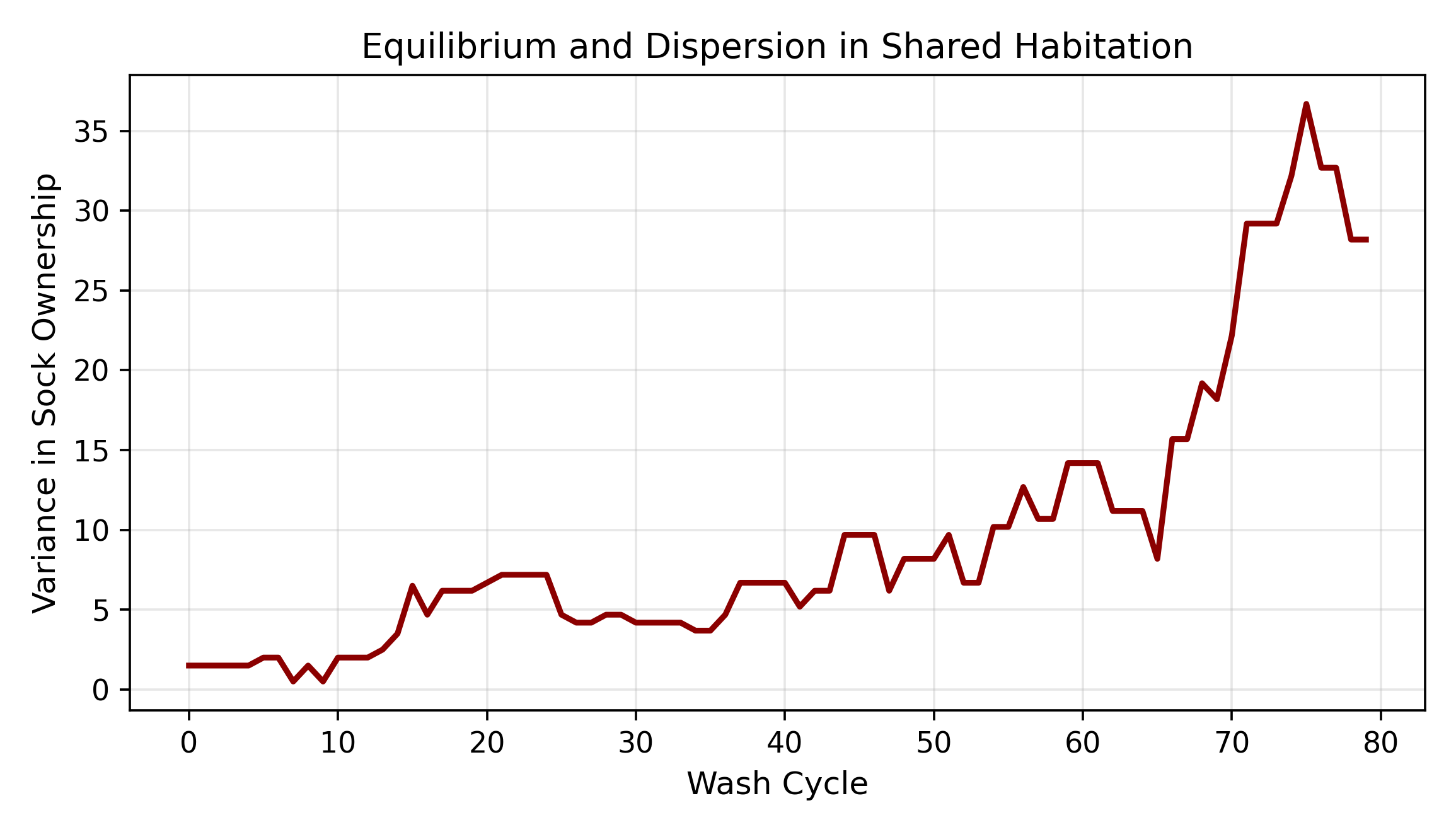}
    \caption{Fluctuations of variance in the number of socks held by 2 cohabitant.}
    \label{fig:sock_exchange_variance}

\end{figure}

\begin{figure}[H]
    \centering
    \includegraphics[width=0.8\textwidth]{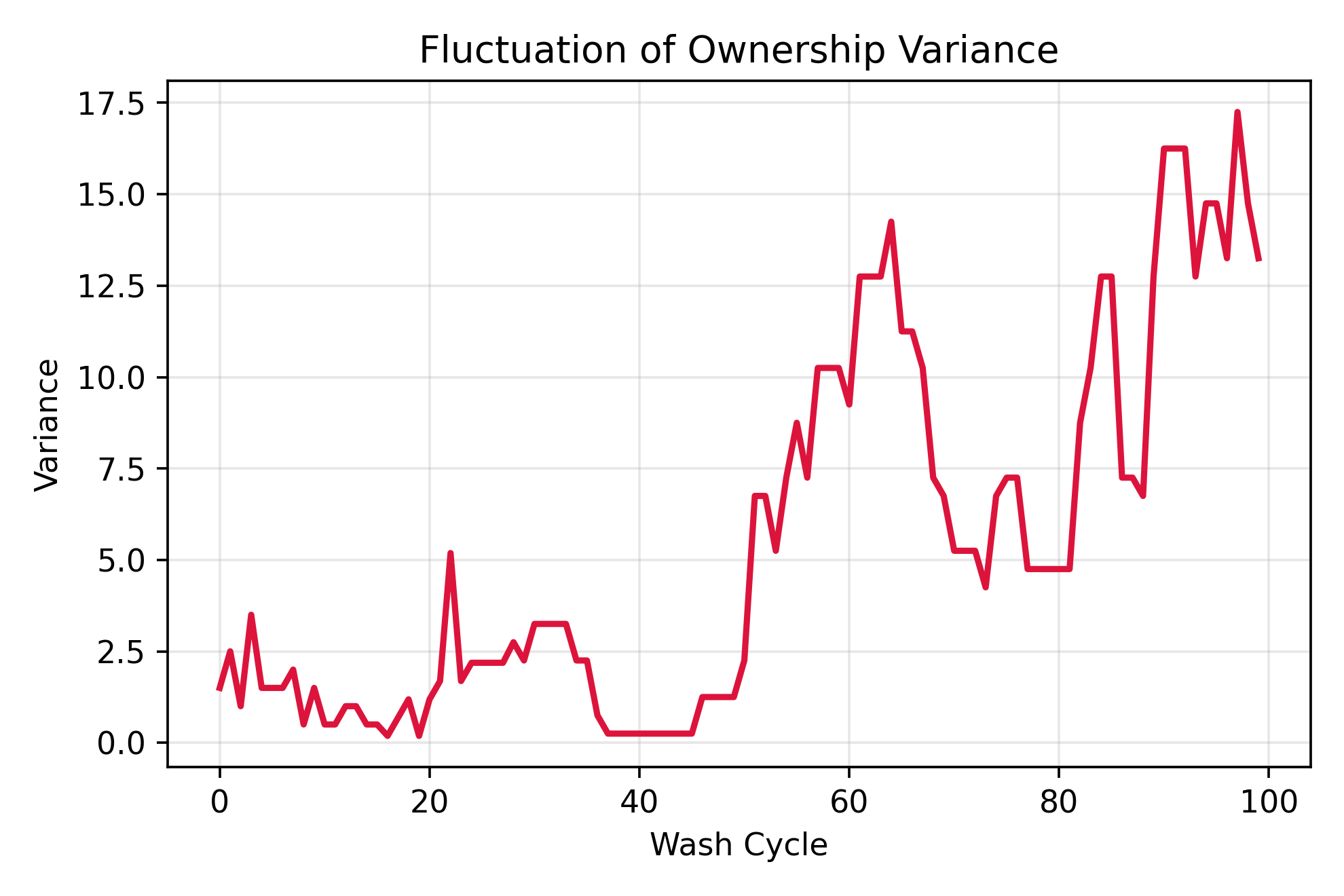}
    \caption{Fluctuations of variance in the number of socks held by N cohabitant.}
    \label{fig:sock_variance_curve}

\end{figure}

\noindent
Figures~\ref{fig:sock_exchange_variance} and~\ref{fig:sock_variance_curve} illustrate the temporal evolution of variance in the number of socks held by two and multiple cohabitants, respectively. Initially, the variance is low, reflecting an almost uniform distribution of items. Over successive laundry cycles, random exchanges, losses, and recoveries increase system disorder, driving the variance upward. After roughly 80 cycles, both systems converge toward stable entropy conditions, where the variance oscillates around a steady equilibrium value.

\subsection{Sock Trajectories}
\begin{figure}[H]
    \centering
    \includegraphics[width=0.8\textwidth]{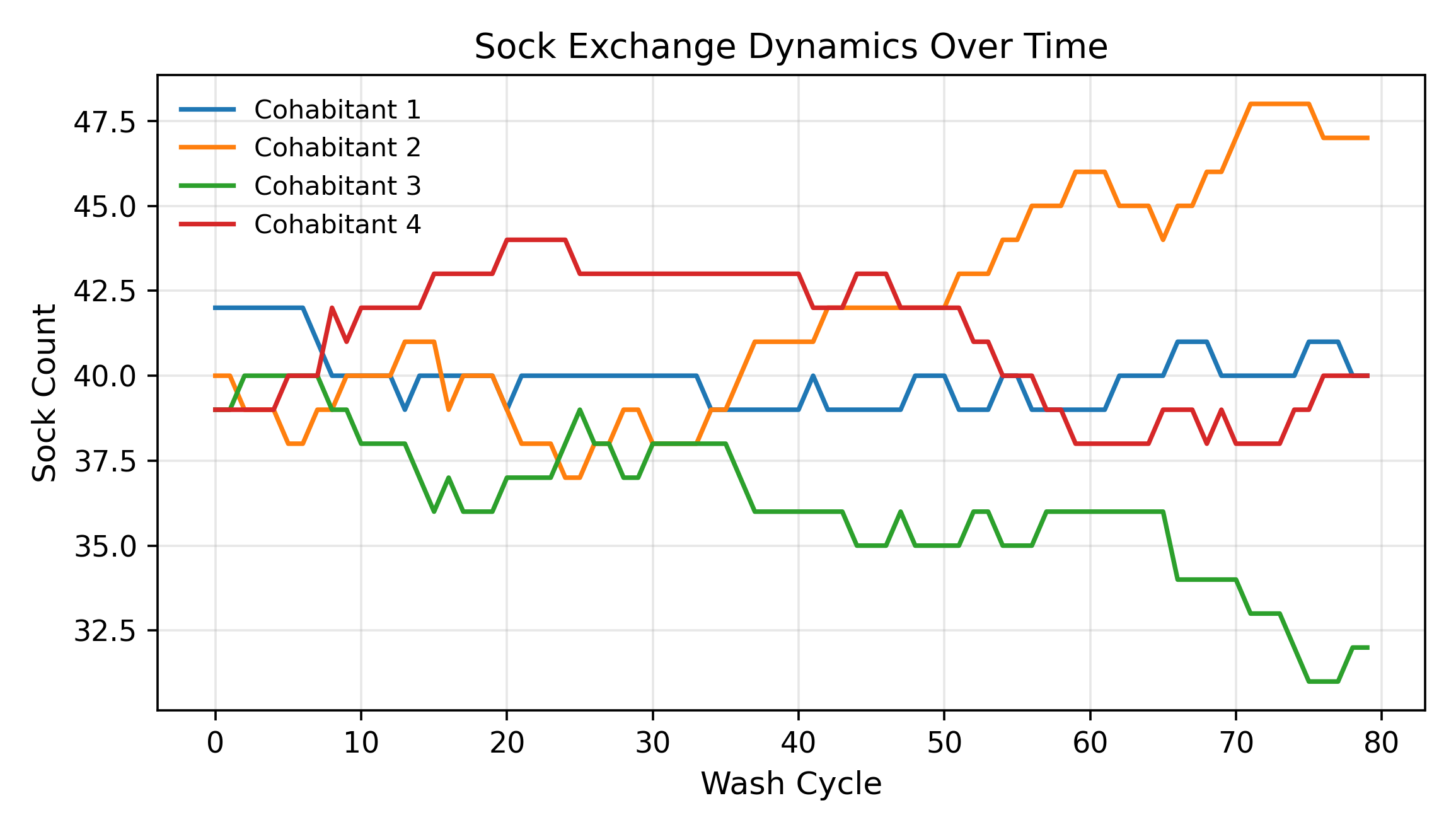}
    \caption{Trajectories of individual sock holdings over time. Each line represents a cohabitant's sock count, illustrating the dynamic nature of the exchange process.}
    \label{fig:sock_exchange_trajectories}

\end{figure}

\begin{figure}[H]
    \centering
    \includegraphics[width=0.8\textwidth]{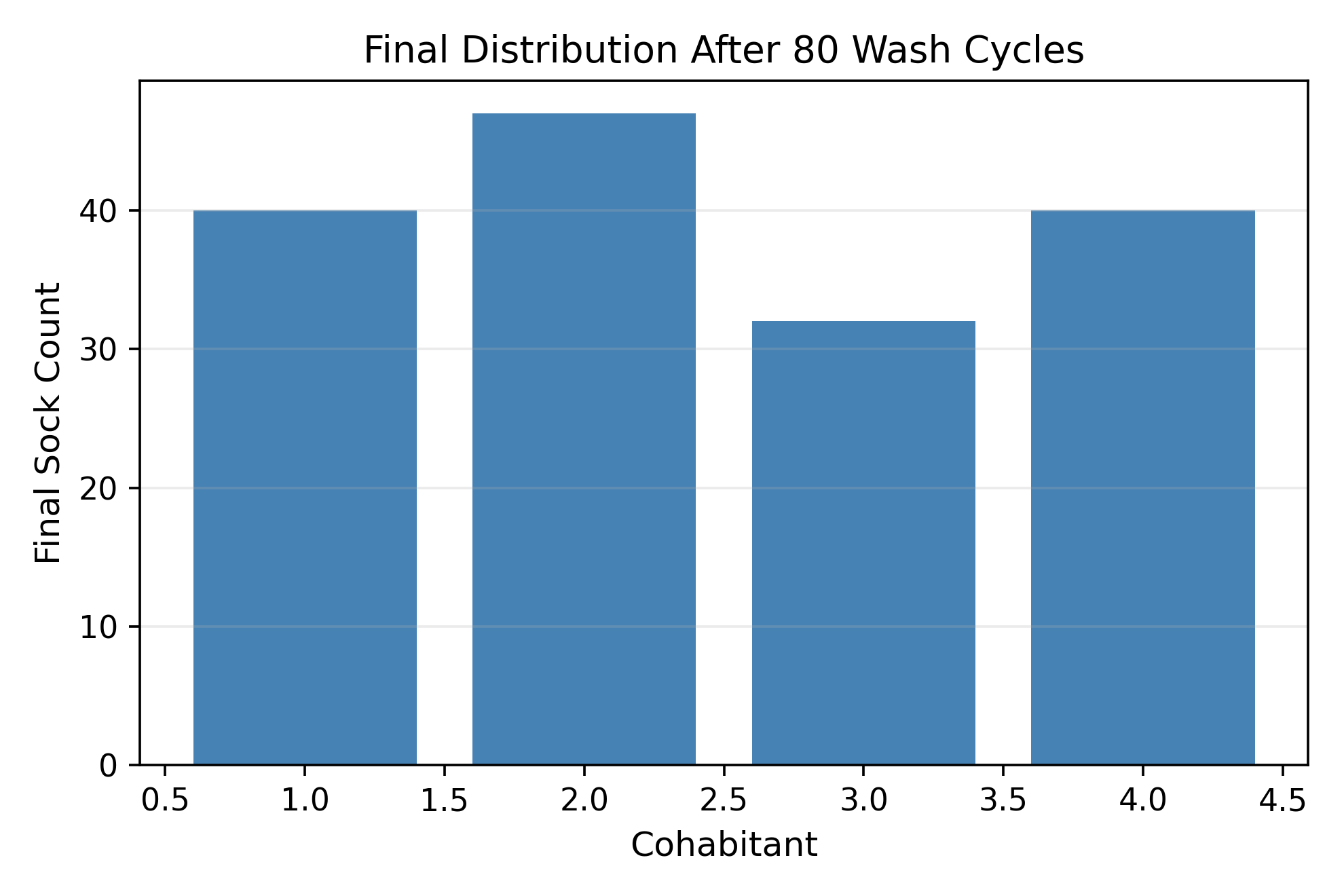}
    \caption{Final distribution of sock holdings among cohabitants. Each bar represents the total number of socks held by a cohabitant at the end of the simulation.}
    \label{fig:sock_exchange_final_distribution}

\end{figure}

\noindent
As shown in Figures~\ref{fig:sock_exchange_trajectories} and~\ref{fig:sock_exchange_final_distribution}, cohabitants with lower exchange probabilities tend to lose socks over successive cycles, while those with higher exchange rates gradually accumulate them. Participants with intermediate exchange rates generally approach a quasi-equilibrium state, maintaining moderate fluctuations around a stable mean. After approximately 80 cycles, the final distribution of sock holdings remains largely unchanged when exchange probabilities are similar across agents; otherwise, persistent asymmetries emerge, reflecting interconnected biases within the stochastic exchange network.

\subsection{Sensitivity Analysis}
\begin{figure}[H]
    \centering
    \includegraphics[width=0.8\textwidth]{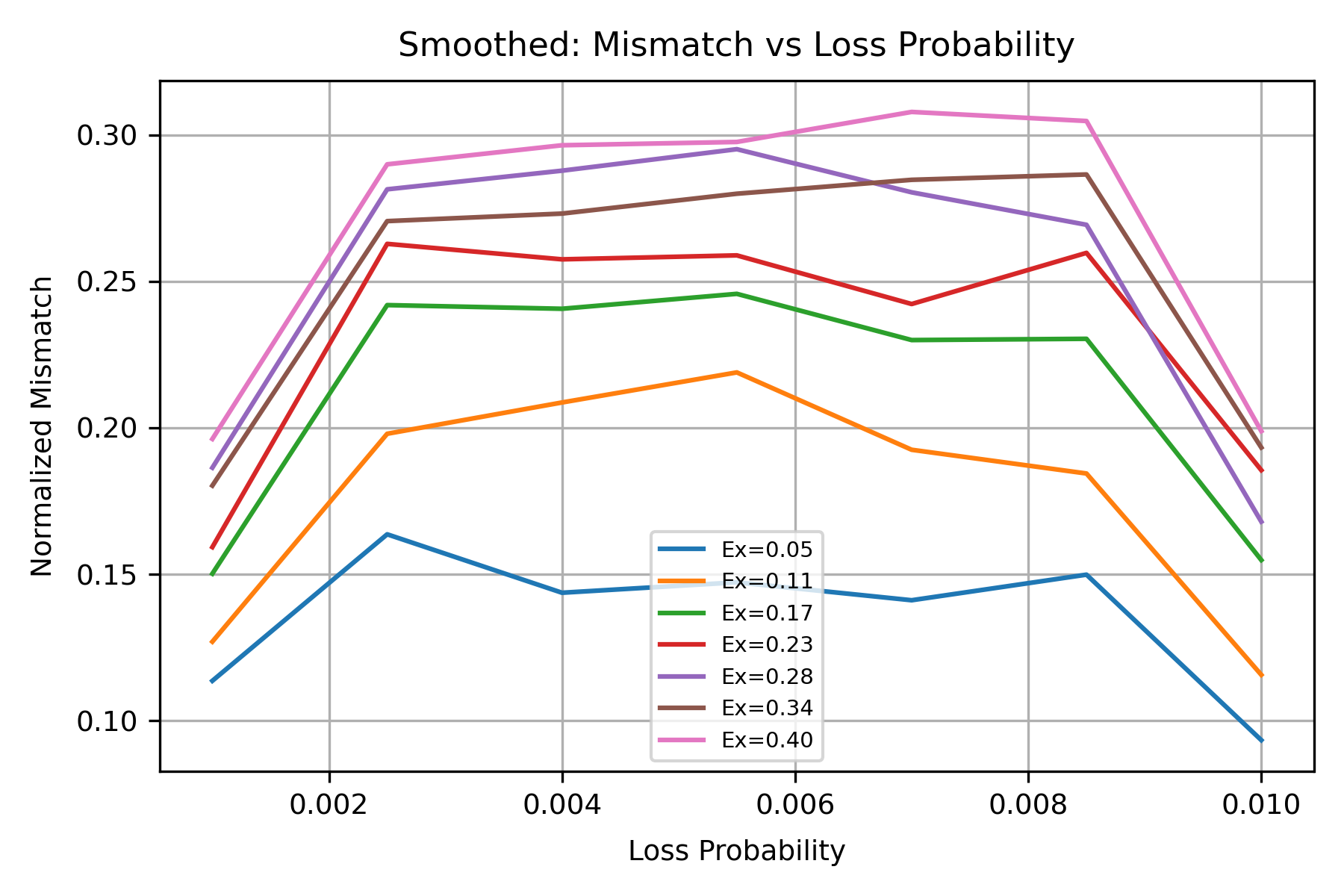}
    \caption{Sensitivity analysis of sock holdings with varying loss probabilities. }
    \label{fig:sockstudy_4_loss_slices}

\end{figure}

\begin{figure}[H]
    \centering
    \includegraphics[width=0.8\textwidth]{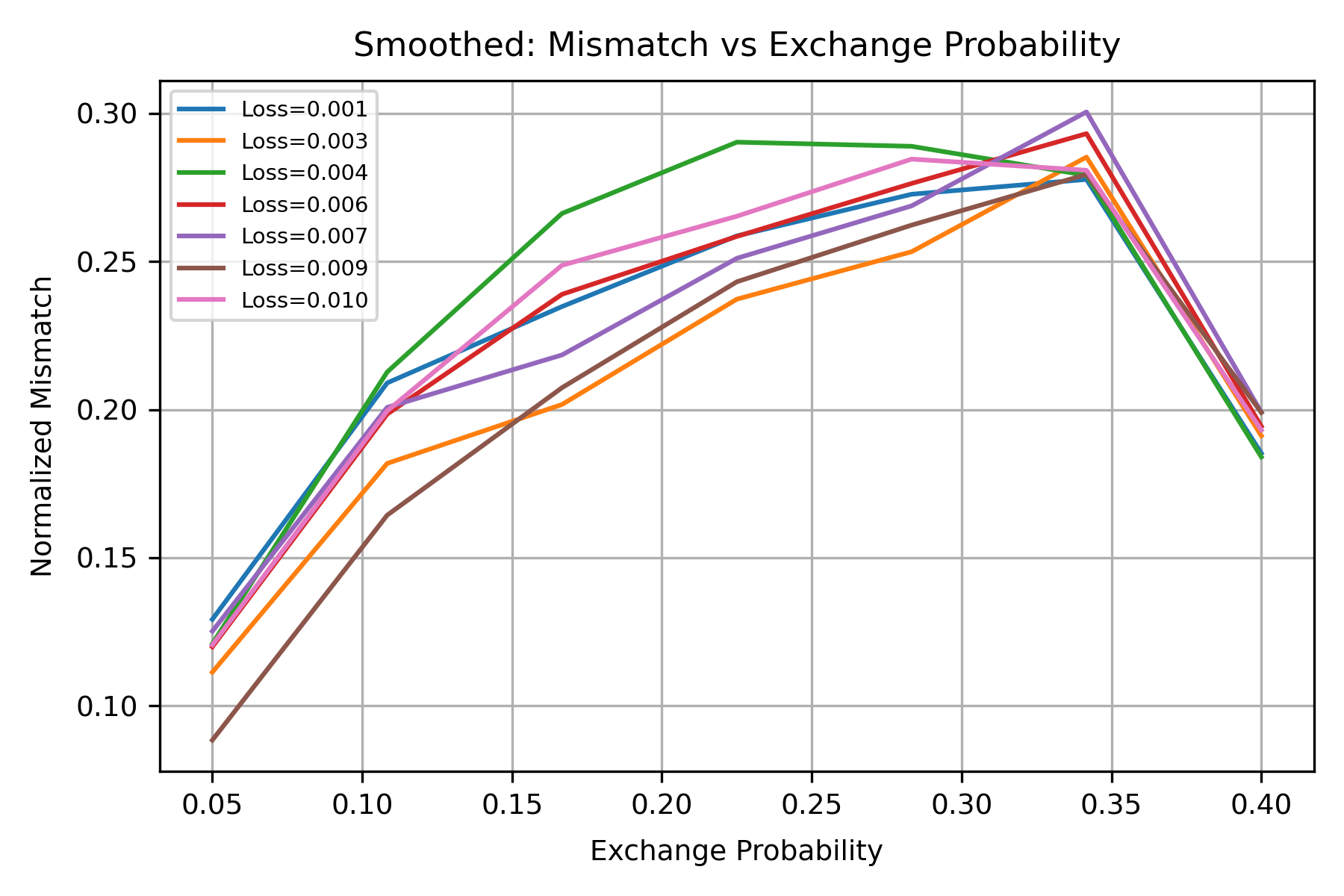}
    \caption{Sensitivity analysis of sock holdings with varying exchange probabilities. }
    \label{fig:sockstudy_5_exchange_slices}

\end{figure}

\noindent
Figures~\ref{fig:sockstudy_4_loss_slices} and~\ref{fig:sockstudy_5_exchange_slices} present the sensitivity analysis of the system under varying loss and exchange probabilities, respectively. In Figure~\ref{fig:sockstudy_5_exchange_slices}, the exchange probability of approximately 0.35 produces the maximum mismatch among cohabitants, while deviations toward either lower or higher exchange rates reduce this disparity. For the loss parameter in Figure~\ref{fig:sockstudy_4_loss_slices}, the mismatch remains nearly constant across the 0.004–0.008 range, suggesting a flat response where moderate loss probabilities yield comparable equilibrium behaviors. The highest observed mismatch corresponds to a loss probability near 0.009, highlighting the nonlinear interplay between exchange and attrition in the redistribution process.

\begin{figure}[H]
    \centering
    \includegraphics[width=0.8\textwidth]{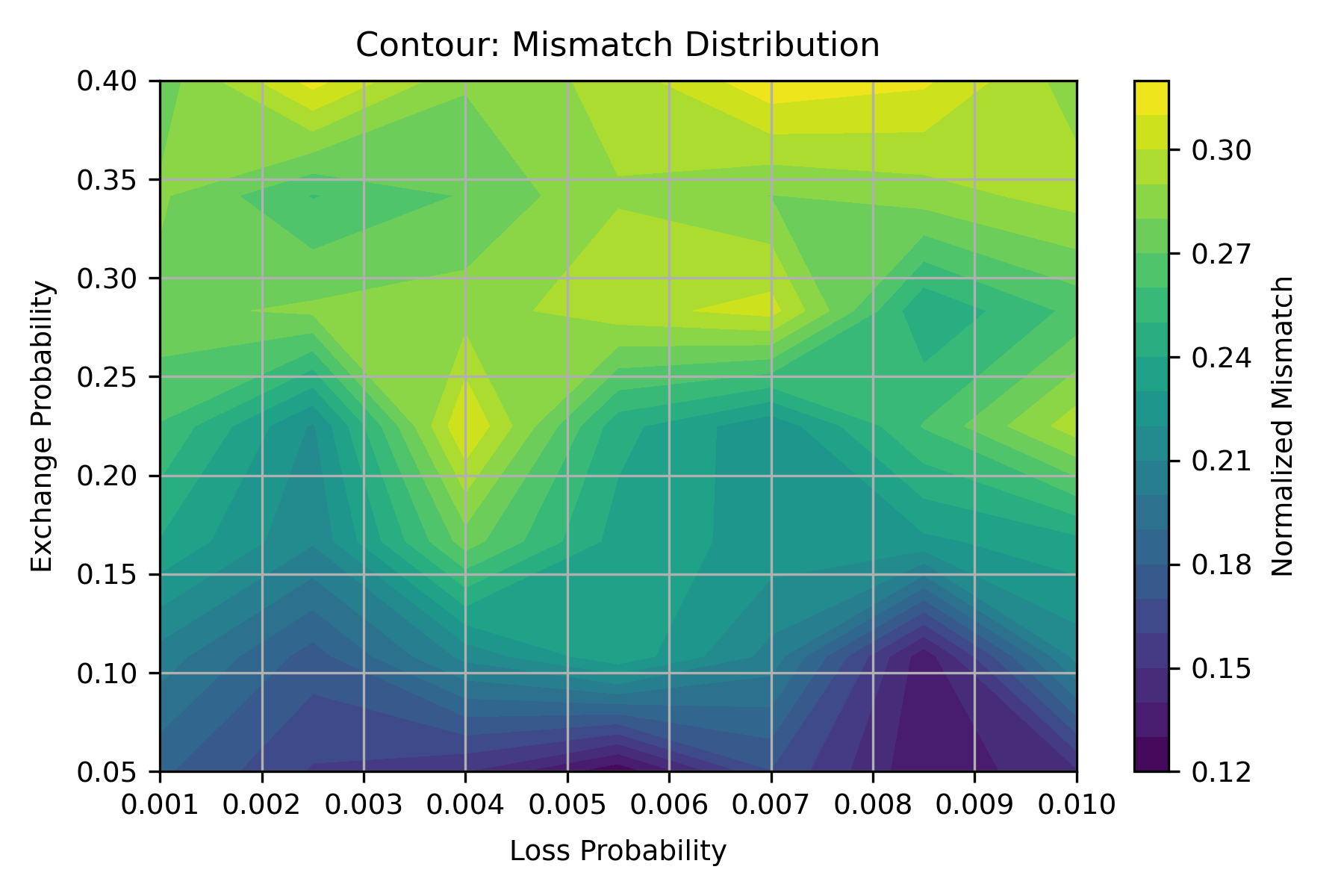}
    \caption{Contour plot showing the relationship between exchange probability, recovery probability, and variance in sock holdings.}
    \label{fig:sockstudy_3_contour}

\end{figure}

\noindent
Figure~\ref{fig:sockstudy_3_contour} illustrates the contour relationship between exchange probability, recovery probability, and the resulting variance in sock holdings. The plot reveals a nonlinear interaction surface where variance peaks occur within a narrow intermediate band of exchange probabilities. At very low exchange rates, agents remain largely isolated, limiting redistribution and producing minimal variance. Conversely, at very high exchange rates, rapid mixing drives the system toward homogenization, again reducing variance. Maximum disorder emerges when moderate exchange coincides with incomplete recovery, reflecting the balance between mixing and selective recollection that sustains long-term variability in the communal system.

\subsection{Analysis}

Figure~\ref{fig:sockstudy_3_contour} complements these observations by visualizing how the interaction between exchange and recovery probabilities shapes system variance. The contour landscape indicates three primary behavioral regions. When \(p_e^U \ll p_r^U\), the system settles into a stable regime where individual ownership remains consistent and recovery dominates mixing. When \(p_e^U \approx p_r^U\), cyclical imbalance emerges: socks oscillate among cohabitants in quasi-periodic fashion, producing alternating peaks and troughs of mismatch similar to those seen in Figures~\ref{fig:sockstudy_4_loss_slices} and~\ref{fig:sockstudy_5_exchange_slices}. Finally, when \(p_e^U > p_r^U\), the model enters a diffusive regime in which items spread randomly through the population, maximizing entropy and mismatch. Across all states, the loss parameters \(p_l^I\) and \(p_l^U\) introduce gradual exponential decay, ensuring that even stable systems experience long-term reduction in total item count, consistent with the entropy growth trend identified in the sensitivity analyses.

\section{Discussion}
\paragraph{}

The model exhibits several emergent properties shaped by the coupled influence of exchange, correction, and loss dynamics. At low exchange probabilities, the system stabilizes into an equilibrium configuration where each agent retains an almost constant proportion of foreign socks, indicating efficient yet contained redistribution. This regime mirrors conditions of high self-recognition and low diffusion, where corrective mechanisms dominate over random mixing. As the exchange probability increases to moderate levels, asymmetry begins to appear: some agents progressively accumulate disproportionate shares of others’ belongings, leading to sustained imbalance. This transition highlights the sensitivity of the system to stochastic variations in exchange events, where minor probabilistic biases compound into long-term inequities. Entropy, representing system disorder, exhibits near-logarithmic growth before reaching saturation, balancing random exchange and corrective recovery. Over extended cycles, cumulative losses follow an exponential decay pattern consistent with attrition dynamics observed in ecological systems.

\subsection{Physical Analogies}
The interplay between disorder growth and corrective stabilization resembles thermodynamic equilibrium in dissipative systems. Exchange events act as microstates driving entropy, while correction introduces a restorative counterforce akin to negative feedback in closed systems. The saturation of disorder parallels the steady-state entropy plateau of physical systems constrained by finite energy and dissipation limits.

\subsection{Social Analogies}
Within a social context, the simulation evokes parallels with resource drift, ownership imbalance, and unequal redistribution in shared environments. Even under uniform probabilistic rules, stochastic noise produces persistent inequality, illustrating how randomness alone can generate structural asymmetry. The corrective process, while stabilizing, remains limited by imperfect feedback, echoing real-world constraints of perception, memory, and social cooperation.

\subsection{System-Theoretic Implications}
From a systems perspective, the model exemplifies the nonlinear interplay between order and disorder in decentralized environments. Exchange and correction operate as competing feedback loops with one amplifying entropy, the other restoring equilibrium. The system’s emergent steady state demonstrates how complex collective behavior can arise from simple local rules, underscoring the inherent fragility of equilibrium when recovery capacity is finite. These dynamics highlight a general principle of self-organizing systems: stability is conditional, temporary, and perpetually negotiated against loss.

\section{Conclusion}
\paragraph{}
\noindent
Viewed through the lens of an ordinary domestic activity, this study demonstrates that the redistribution of indistinguishable objects among cohabitants arises naturally from stochastic exchange dynamics. The discrete-event model captures how small-scale random interactions can amplify into large-scale disorder, echoing the self-organizing principles observed in physical, ecological, and social systems. The results underscore that even in the absence of coordination or intent, collective behaviors tend to converge toward quasi-equilibrium configurations defined by the probabilistic balance of exchange, recovery, and loss. This emergent equilibrium is inherently dynamic, characterized by continual fluctuations that preserve systemic entropy within bounded limits.  

Beyond the humorous context of laundry systems, these findings reveal general insights into how randomness and limited feedback control shape redistributive processes. Analogous behaviors may be found in domains ranging from supply-chain inefficiencies and shared digital resource allocation to ecological population drift and economic exchange models. The persistence of imbalance under symmetric probabilistic rules reinforces the notion that fairness and stability in distributed systems are emergent properties rather than guaranteed outcomes.

However, several limitations constrain the scope of the present model. The simulation assumes homogeneous agents with identical behavioral parameters, neglecting heterogeneity in habits, perception accuracy, and recovery efficiency that characterize real-world interactions. External factors such as cooperative intent, communication, or hierarchical influence were not incorporated, potentially oversimplifying the collective dynamics. Moreover, the temporal horizon of the simulation, while sufficient to capture steady-state behavior, does not fully explore long-term evolutionary adaptation or the impact of stochastic

\bibliographystyle{ieeetr} 
\bibliography{sock}

\end{document}